\documentclass[%
 reprint,
 superscriptaddress,
nofootinbib,
 amsmath,amssymb,
 aps,
]{revtex4-2}

\usepackage{graphicx}% Include figure files
\usepackage{dcolumn}% Align table columns on decimal point
\usepackage{bm}% bold math
\usepackage{todonotes}
\usepackage[many]{tcolorbox}
\usepackage{lipsum}
\usepackage{amsmath,amssymb,cancel}

\usepackage{shuffle}
\usepackage{graphbox}
\usepackage{environ}
\usepackage{physics}
\usepackage{tabularx}
\usepackage{MnSymbol}
\usetikzlibrary{shapes.geometric}
\usepackage{asymptote}
\usepackage{float}
\usepackage{enumitem}
\usepackage{tikz,graphicx,adjustbox}
\newcommand*{\shifttext}[2]{%
  \settowidth{\@tempdima}{#2}%
  \makebox[\@tempdima]{\hspace*{#1}#2}%
}

\usepackage[T1]{fontenc}
\usepackage{amsmath,amssymb,bm,tikz,hyperref}
\usetikzlibrary{arrows.meta}
\usepackage{amsmath,tikz}

\newcommand{\BGraph}[5]{%
\vcenter{\hbox{%
\begin{tikzpicture}[x=.9cm,line width=.8pt]
  \node[circle,draw,fill=black,inner sep=2.5pt,
        label={[font=\small]above:$a$},
        label={[font=\scriptsize]below:{$#1$}}]
        (a) at (0,0) {};
  \node[circle,draw,fill=white,inner sep=2.5pt,
        label={[font=\small]above:$b$},
        label={[font=\scriptsize]below:{$#2$}}]
        (b) at (1.1,0) {};
  \node[circle,draw,fill=black,inner sep=2.5pt,
        label={[font=\small]above:$c$},
        label={[font=\scriptsize]below:{$#3$}}]
        (c) at (2.2,0) {};
  \ifnum#4=1\relax \draw (a)--(b); \fi
  \ifnum#5=1\relax \draw (b)--(c); \fi
\end{tikzpicture}%
}}%
}
\hypersetup{colorlinks=true,linkcolor=blue,citecolor=blue,urlcolor=blue}

\newcommand{\Se}{\langle 23\rangle\langle 45\rangle\langle 61\rangle}
\newcommand{\So}{\langle 12\rangle\langle 34\rangle\langle 56\rangle}

\usepackage{xcolor}
\definecolor{kyblue}{RGB}{70,160,220}

\begin{document}

\noindent\hfill MPP-2026-203
\vspace{-2ex} % 把標題往上拉回來一點（可調整）

\title{Simplicity of the Three-Loop Six-Point ABJM Amplitude}

\author{Hanyu Fang}
\email{aurora\_superposition@sjtu.edu.cn}
\affiliation{School of Physics and Astronomy, Shanghai Jiao Tong University, Shanghai 200240, China}
\affiliation{Key Laboratory for Particle Astrophysics and Cosmology (MOE), Shanghai 200240, China}

\author{Chia-Kai Kuo}
\email{chia-kai.kuo@mpp.mpg.de}
\affiliation{Max–Planck–Institut f\"ur Physik, Werner–Heisenberg–Institut, D–85748 Garching bei M\"unchen, Germany}

\author{Kai Yan}%
\email{yan.kai@sjtu.edu.cn}
\affiliation{School of Physics and Astronomy, Shanghai Jiao Tong University, Shanghai 200240, China}
\affiliation{Key Laboratory for Particle Astrophysics and Cosmology (MOE), Shanghai 200240, China}

%We compute the three-loop six-point amplitude in planar ABJM theory using negative geometries. At three loops, only a single bipartite chain contributes to the logarithm of the amplitude. Remarkably, integrating out its two endpoint loops already yields $-2\pi^2$ times the one-loop integrand, so the full integral is infrared finite, equal to $-\pi^2$ times the one-loop amplitude. Hence the collinear anomalous dimension, like the cusp, has no three-loop correction. Restoring the disconnected products of lower-loop contributions gives the full three-loop amplitude, whose infrared divergences arise entirely from these products. Although it has uniform transcendental weight three, the amplitude involves only $\pi$ times weight-two functions already present at two loops. These results reveal an unexpected simplicity of the three-loop ABJM amplitude.

\begin{abstract}
We compute the three-loop six-point amplitude in planar ABJM theory using negative geometries. At this order, only a single bipartite chain contributes to the logarithm of the amplitude, and we find that its dual-conformally invariant integrand is locally finite. The integrated chain is therefore infrared finite and, remarkably, equals exactly $-\pi^2$ times the one-loop amplitude. Hence the collinear anomalous dimension, like the cusp, receives no three-loop correction. Restoring the disconnected products of lower-loop contributions, which carry all the infrared divergences, we obtain the full three-loop amplitude, fixed entirely by lower-loop data and the coefficient $-\pi^2$. These results reveal an unexpected simplicity of the six-point ABJM amplitude and suggest an iterative relation between its odd- and even-loop terms.

%\ky{
%Nontrivial dual-conformal cross-ratios first appear at six points in planar ABJM theory. We construct and evaluate the three-loop six-point amplitude using negative geometries. The negative-geometry expansion organizes the logarithm of the amplitude into connected bipartite graphs, leaving a single chain topology with a dual-conformally invariant, locally finite integrand. In this formalism we uncover unexpected simplicity:  the integrated three-loop negative geometry  equals exactly $-\pi^2$  times the one-loop amplitude.  The complete three-loop amplitude is thus determined by lower-loop data and a single new constant. This exact relation proves the vanishing of the three-loop correction to the cusp anomalous dimension and sheds light on an iterative structure for six-point ABJM amplitudes. }
\end{abstract}

\maketitle

\section{Introduction}

Scattering amplitudes in three-dimensional $\mathcal N=6$ superconformal Chern--Simons--matter (ABJM) theory~\cite{Aharony:2008ug,Hosomichi:2008jb} share remarkable structures with planar $\mathcal N=4$ super-Yang--Mills theory, including Yangian symmetry and Grassmannian representations~\cite{Bargheer:2010hn,Huang:2010qy,Arkani-Hamed:2012zlh,Gang:2010gy,Huang:2013owa,Huang:2014xza,Lee:2010du}. These parallels extend to geometric descriptions of amplitudes~\cite{Arkani-Hamed:2013jha,He:2022cup,He:2023rou,Damgaard:2019ztj,Arkani-Hamed:2017mur,He:2021llb,Huang:2021jlh,Lukowski:2023nnf,Ferro:2025pij,He:2025zen,He:2026ibk}. In particular, the ABJM amplituhedron is obtained by a symplectic reduction of its four-dimensional counterpart, together with a reversal of the relevant sign conditions~\cite{He:2022cup,He:2023rou}. Its canonical forms encode the loop integrands, and the negative-geometry expansion rewrites them as sums over graphs; the connected graphs give the integrand of the logarithm of the amplitude~\cite{Arkani-Hamed:2021iya}. In ABJM theory this expansion further simplifies to bipartite graphs~\cite{He:2022cup,He:2023rou}. At four points, integrating out all but one loop variable produces infrared-finite functions related to the cusp anomalous dimension~\cite{He:2023exb,Henn:2023pkc,Lagares:2024epo,Li:2024lbw}. The amplitude itself, however, requires integrating out all $L$ loop variables.

Perturbative studies of ABJM amplitudes have made substantial progress, from one-loop amplitudes at arbitrary multiplicity~\cite{Brandhuber:2012wy} to two-loop amplitudes at four, six, and eight points~\cite{Bianchi:2011dg,Caron-Huot:2012sos,He:2022lfz} and the two-loop infrared structure at arbitrary multiplicity~\cite{Huang:2025hya}. Beyond two loops, the only integrated result is the three-loop four-point amplitude, which provides evidence for exponentiation including finite terms~\cite{Bianchi:2014iia}. At four points, however, there are no dual-conformal cross-ratios. Six points is the first multiplicity at which they appear, and the amplitude involves two independent Yangian invariants; at two loops, one multiplies a nontrivial remainder function and the other an additional finite function~\cite{Caron-Huot:2012sos}.
The three-loop six-point amplitude is therefore the natural place to ask whether new functions of the cross-ratios appear beyond two loops.
\iffalse
The three-loop six-point amplitude is therefore a natural next step, which also tests whether the simplicity of negative geometries survives full loop integration.
\fi

In this Letter, we compute the three-loop six-point ABJM amplitude by organizing its integrand into negative geometries and integrating the connected part over all three loop variables. Bipartiteness restricts the connected three-loop topology to a chain, whose numerators are fixed by matching two-particle unitarity cuts.   Remarkably, the connected three-loop integrand is
manifestly dual conformally invariant (DCI) and locally finite,
allowing direct integration in three dimensions without
an infrared regulator.  Therefore \emph{the three-loop connected negative geometry remains infrared finite even after all three loop variables have been integrated out}, extending the finiteness previously established for partially integrated negative geometries. The result takes a simple form, proportional to $\pi^3$. This simplicity is already visible before the last loop integration: integrating out the two endpoint loops of the chain gives $-2\pi^2$ times the one-loop six-point integrand.

The fully integrated connected negative geometry directly determines the three-loop contribution to the logarithm of the amplitude. Writing $M_\pm$ for the perturbative corrections multiplying the two independent six-point leading singularities, we find
\begin{equation}\label{eq:prl-intro-exponentiation}
\left.\log M_\pm\right|_{\text{three loops}}
=-\pi^2 M_\pm^{(1)}.
\end{equation}
 The full three-loop amplitude is then obtained by restoring the disconnected products of lower-loop contributions, which account for its infrared divergences. 
Equation~\eqref{eq:prl-intro-exponentiation} reveals a hidden simplicity of the three-loop amplitude: no new function of the cross-ratios appears, and up to the constant $-\pi^2$, the amplitude is fixed by the one- and two-loop amplitudes.

    \section{Negative geometry and the three-loop integrand}
    \label{sec:negative-geometry}
    
    We consider the planar, color-ordered six-point superamplitude and strip off a factor $g^L$ at $L$ loops, with $g=N/k$. Following Ref.~\cite{Caron-Huot:2012sos}, we use dual coordinates $p_i=x_{i+1}-x_i$ and embedding vectors $y_i=(x_i,1,x_i^2)$, so that $(i\cdot j)=y_i\cdot y_j=x_{ij}^2$ and $(i\cdot i+1)=0$. The embedding-space $\epsilon$ tensor is normalized by $\epsilon(i_1,\dots,i_5)\,\epsilon(j_1,\dots,j_5)=\det[(i_r\cdot j_s)]$. The integration measure is $\int_a=4\pi\int d^3x_a/[i(2\pi)^3]$. The kinematics depends on the cross-ratios
    \begin{equation}\label{eq:u}
    u_1=\frac{(1\cdot 3)(4\cdot 6)}{(1\cdot 4)(3\cdot 6)},\;\;
    u_2=\frac{(2\cdot 4)(5\cdot 1)}{(2\cdot 5)(4\cdot 1)},\;\;
    u_3=\frac{(3\cdot 5)(6\cdot 2)}{(3\cdot 6)(5\cdot 2)},
    \end{equation}
    subject to the three-dimensional Gram constraint $(1-u_1-u_2-u_3)^2=4u_1u_2u_3$.
    
    The six-point amplitude has two leading singularities~\cite{Huang:2013owa,Huang:2014xza,Caron-Huot:2012sos}, $\mathrm{LS}_{6,\pm}=\frac12[\mathcal A_6^{\mathrm{tree}}\mp i\mathcal A_{6,\mathrm{shifted}}^{\mathrm{tree}}]$, where the shifted tree $\mathcal A_{6,\mathrm{shifted}}^{\mathrm{tree}}=\mathcal A_6^{\mathrm{tree}}(\bar2,3,\bar4,5,\bar6,1)$ is obtained by shifting all labels by one site. At $L$ loops, we write
    \begin{gather}
    \mathcal A_6^{(L)}=\mathrm{LS}_{6,+}M_+^{(L)}+\mathrm{LS}_{6,-}M_-^{(L)},\notag\\
    M_\pm^{(L)}=\frac1{L!}\int_{a_1,\dots,a_L}\Omega^{(L)}_{6,\pm}.\label{eq:LS-decomposition}
    \end{gather}
    Here $\Omega^{(L)}_{6,\pm}$ denotes the loop integrand with the measure stripped off. On the positive branch it is the canonical form of the six-point ABJM amplituhedron~\cite{He:2023rou}; the negative branch follows from $\lambda_{1,3,5}\mapsto-\lambda_{1,3,5}$, which leaves the dual coordinates and the embedding-space $\epsilon$ tensors unchanged but flips the signs of the spinor products $\langle 12\rangle\langle 34\rangle\langle 56\rangle$ and $\langle 23\rangle\langle 45\rangle\langle 61\rangle$. We therefore focus on $\Omega^{(L)}_{6,+}$.
    
    \emph{Negative geometries.}---The ABJM amplituhedron imposes the mutual condition $\langle(AB)_a(AB)_b\rangle<0$ on each pair of loop lines. Writing each of these conditions as the unconstrained region minus its complement expands the loop integrand into a signed sum over graphs~\cite{Arkani-Hamed:2021iya,He:2022cup,He:2023rou}. Each vertex represents a loop line $(AB)_a$, each edge $(a,b)$ imposes the reversed condition $\langle(AB)_a(AB)_b\rangle>0$, and a graph with $E$ edges enters with the sign $(-1)^E$. Suppressing loop labels and multiplicities, the two- and three-loop expansions read
    \begin{equation}\label{eq:ng-expansion}
    \adjustbox{max width=0.97\columnwidth}{%
      $\displaystyle \vcenter{\hbox{\scalebox{0.8}{%
    \begin{tikzpicture}[x=0.75pt,y=0.75pt,yscale=-1,xscale=1]
    % Three-loop triangle graph
    \draw  [line width=0.5]  (304.41,152.52) -- (329.68,195.77) -- (279.13,195.77) -- cycle ;
    \draw  [fill={rgb, 255:red, 155; green, 155; blue, 155 }  ,fill opacity=1 ][line width=0.5]  (275.03,195.77) .. controls (275.03,193.5) and (276.87,191.67) .. (279.13,191.67) .. controls (281.4,191.67) and (283.23,193.5) .. (283.23,195.77) .. controls (283.23,198.03) and (281.4,199.87) .. (279.13,199.87) .. controls (276.87,199.87) and (275.03,198.03) .. (275.03,195.77) -- cycle ;
    \draw  [fill={rgb, 255:red, 155; green, 155; blue, 155 }  ,fill opacity=1 ][line width=0.5]  (325.58,195.77) .. controls (325.58,193.5) and (327.42,191.67) .. (329.68,191.67) .. controls (331.95,191.67) and (333.78,193.5) .. (333.78,195.77) .. controls (333.78,198.03) and (331.95,199.87) .. (329.68,199.87) .. controls (327.42,199.87) and (325.58,198.03) .. (325.58,195.77) -- cycle ;
    \draw  [fill={rgb, 255:red, 155; green, 155; blue, 155 }  ,fill opacity=1 ][line width=0.5]  (300.31,152.52) .. controls (300.31,150.26) and (302.14,148.42) .. (304.41,148.42) .. controls (306.67,148.42) and (308.51,150.26) .. (308.51,152.52) .. controls (308.51,154.79) and (306.67,156.62) .. (304.41,156.62) .. controls (302.14,156.62) and (300.31,154.79) .. (300.31,152.52) -- cycle ;
    % Two-loop edge
    \draw   [line width=0.5] (264.81,88.86) -- (315.36,88.86) ;
    \draw  [line width=0.5][fill={rgb, 255:red, 155; green, 155; blue, 155 }  ,fill opacity=1 ] (311.26,88.86) .. controls (311.26,86.59) and (313.09,84.76) .. (315.36,84.76) .. controls (317.62,84.76) and (319.46,86.59) .. (319.46,88.86) .. controls (319.46,91.12) and (317.62,92.96) .. (315.36,92.96) .. controls (313.09,92.96) and (311.26,91.12) .. (311.26,88.86) -- cycle ;
    \draw [line width=0.5] [fill={rgb, 255:red, 155; green, 155; blue, 155 }  ,fill opacity=1 ] (260.71,88.86) .. controls (260.71,86.59) and (262.54,84.76) .. (264.81,84.76) .. controls (267.07,84.76) and (268.91,86.59) .. (268.91,88.86) .. controls (268.91,91.12) and (267.07,92.96) .. (264.81,92.96) .. controls (262.54,92.96) and (260.71,91.12) .. (260.71,88.86) -- cycle ;
    % Two-loop disconnected pair
    \draw  [line width=0.5][fill={rgb, 255:red, 155; green, 155; blue, 155 }  ,fill opacity=1 ] (425.46,89.26) .. controls (425.46,86.99) and (427.29,85.16) .. (429.56,85.16) .. controls (431.82,85.16) and (433.66,86.99) .. (433.66,89.26) .. controls (433.66,91.52) and (431.82,93.36) .. (429.56,93.36) .. controls (427.29,93.36) and (425.46,91.52) .. (425.46,89.26) -- cycle ;
    \draw  [line width=0.5][fill={rgb, 255:red, 155; green, 155; blue, 155 }  ,fill opacity=1 ] (374.91,89.26) .. controls (374.91,86.99) and (376.74,85.16) .. (379.01,85.16) .. controls (381.27,85.16) and (383.11,86.99) .. (383.11,89.26) .. controls (383.11,91.52) and (381.27,93.36) .. (379.01,93.36) .. controls (376.74,93.36) and (374.91,91.52) .. (374.91,89.26) -- cycle ;
    % Three-loop chain
    \draw    [line width=0.5](181.14,174.86) -- (231.69,174.86) ;
    \draw [line width=0.5] [fill={rgb, 255:red, 155; green, 155; blue, 155 }  ,fill opacity=1 ] (227.59,174.86) .. controls (227.59,172.59) and (229.43,170.76) .. (231.69,170.76) .. controls (233.96,170.76) and (235.79,172.59) .. (235.79,174.86) .. controls (235.79,177.12) and (233.96,178.96) .. (231.69,178.96) .. controls (229.43,178.96) and (227.59,177.12) .. (227.59,174.86) -- cycle ;
    \draw  [line width=0.5][fill={rgb, 255:red, 155; green, 155; blue, 155 }  ,fill opacity=1 ] (177.04,174.86) .. controls (177.04,172.59) and (178.88,170.76) .. (181.14,170.76) .. controls (183.41,170.76) and (185.24,172.59) .. (185.24,174.86) .. controls (185.24,177.12) and (183.41,178.96) .. (181.14,178.96) .. controls (178.88,178.96) and (177.04,177.12) .. (177.04,174.86) -- cycle ;
    \draw [line width=0.5] [fill={rgb, 255:red, 155; green, 155; blue, 155 }  ,fill opacity=1 ] (202.42,174.86) .. controls (202.42,172.59) and (204.25,170.76) .. (206.52,170.76) .. controls (208.78,170.76) and (210.62,172.59) .. (210.62,174.86) .. controls (210.62,177.12) and (208.78,178.96) .. (206.52,178.96) .. controls (204.25,178.96) and (202.42,177.12) .. (202.42,174.86) -- cycle ;
    % Three-loop disconnected triple
    \draw  [line width=0.5][fill={rgb, 255:red, 155; green, 155; blue, 155 }  ,fill opacity=1 ] (422.56,174.16) .. controls (422.56,171.89) and (424.39,170.06) .. (426.66,170.06) .. controls (428.92,170.06) and (430.76,171.89) .. (430.76,174.16) .. controls (430.76,176.42) and (428.92,178.26) .. (426.66,178.26) .. controls (424.39,178.26) and (422.56,176.42) .. (422.56,174.16) -- cycle ;
    \draw  [line width=0.5][fill={rgb, 255:red, 155; green, 155; blue, 155 }  ,fill opacity=1 ] (372.01,174.16) .. controls (372.01,171.89) and (373.84,170.06) .. (376.11,170.06) .. controls (378.37,170.06) and (380.21,171.89) .. (380.21,174.16) .. controls (380.21,176.42) and (378.37,178.26) .. (376.11,178.26) .. controls (373.84,178.26) and (372.01,176.42) .. (372.01,174.16) -- cycle ;
    \draw [line width=0.5] [fill={rgb, 255:red, 155; green, 155; blue, 155 }  ,fill opacity=1 ] (397.38,174.16) .. controls (397.38,171.89) and (399.22,170.06) .. (401.48,170.06) .. controls (403.75,170.06) and (405.58,171.89) .. (405.58,174.16) .. controls (405.58,176.42) and (403.75,178.26) .. (401.48,178.26) .. controls (399.22,178.26) and (397.38,176.42) .. (397.38,174.16) -- cycle ;
    % Three-loop edge plus isolated vertex
    \draw    [line width=0.5](493.6,173.56) -- (528.13,173.56) ;
    \draw [line width=0.5] [fill={rgb, 255:red, 155; green, 155; blue, 155 }  ,fill opacity=1 ] (524.03,173.56) .. controls (524.03,171.29) and (525.86,169.46) .. (528.13,169.46) .. controls (530.39,169.46) and (532.23,171.29) .. (532.23,173.56) .. controls (532.23,175.82) and (530.39,177.66) .. (528.13,177.66) .. controls (525.86,177.66) and (524.03,175.82) .. (524.03,173.56) -- cycle ;
    \draw [line width=0.5] [fill={rgb, 255:red, 155; green, 155; blue, 155 }  ,fill opacity=1 ] (473.48,173.56) .. controls (473.48,171.29) and (475.31,169.46) .. (477.58,169.46) .. controls (479.84,169.46) and (481.68,171.29) .. (481.68,173.56) .. controls (481.68,175.82) and (479.84,177.66) .. (477.58,177.66) .. controls (475.31,177.66) and (473.48,175.82) .. (473.48,173.56) -- cycle ;
    \draw [line width=0.5] [fill={rgb, 255:red, 155; green, 155; blue, 155 }  ,fill opacity=1 ] (491.85,173.56) .. controls (491.85,171.29) and (493.69,169.46) .. (495.95,169.46) .. controls (498.21,169.46) and (500.05,171.29) .. (500.05,173.56) .. controls (500.05,175.82) and (498.21,177.66) .. (495.95,177.66) .. controls (493.69,177.66) and (491.85,175.82) .. (491.85,173.56) -- cycle ;
    % Brace under the signed connected two-loop term (includes its minus sign)
    \draw  [line width=0.5] (233,101.67) .. controls (233,106.34) and (235.33,108.67) .. (240,108.67) -- (265.5,108.67) .. controls (272.17,108.67) and (275.5,111) .. (275.5,115.67) .. controls (275.5,111) and (278.83,108.67) .. (285.5,108.67)(282.5,108.67) -- (310.67,108.67) .. controls (315.34,108.67) and (317.67,106.34) .. (317.67,101.67) ;
    % Brace under the connected three-loop terms
    \draw  [line width=0.5] (179.33,208.67) .. controls (179.33,213.34) and (181.66,215.67) .. (186.33,215.67) -- (246.17,215.67) .. controls (252.84,215.67) and (256.17,218) .. (256.17,222.67) .. controls (256.17,218) and (259.5,215.67) .. (266.17,215.67)(263.17,215.67) -- (326,215.67) .. controls (330.67,215.67) and (333,213.34) .. (333,208.67) ;
    % Text nodes
    \draw (444,88) node [anchor=north west][inner sep=0.75pt]  [font=\large] [align=left] {$\displaystyle ,$};
    \draw (544,172) node [anchor=north west][inner sep=0.75pt]  [font=\large] [align=left] {$\displaystyle .$};
    \draw (234.98,80.95) node [anchor=north west][inner sep=0.75pt]  [font=\large] [align=left] {$\displaystyle -$};
    \draw (340.18,80.95) node [anchor=north west][inner sep=0.75pt]  [font=\large] [align=left] {$\displaystyle +$};
    \draw (227,78.15) node [anchor=north east][inner sep=0.75pt]  [font=\Large] [align=left] {$\displaystyle \Omega_{6,+}^{(2)}=$};
    \draw (252,165.85) node [anchor=north west][inner sep=0.75pt]  [font=\large] [align=left] {$\displaystyle -$};
    \draw (170,165.85) node [anchor=north east][inner sep=0.75pt]  [font=\Large] [align=left] {$\displaystyle \Omega_{6,+}^{(3)}=$};
    \draw (342,165.85) node [anchor=north west][inner sep=0.75pt]  [font=\large] [align=left] {$\displaystyle +$};
    \draw (443.68,165.85) node [anchor=north west][inner sep=0.75pt]  [font=\large] [align=left] {$\displaystyle -$};
    \draw (275.5,117.48) node [anchor=north][inner sep=0.75pt]  [font=\large] [align=left] {$\displaystyle \widetilde{\Omega}_{6,+}^{(2)}$};
    \draw (256.17,226.48) node [anchor=north][inner sep=0.75pt]  [font=\large] [align=left] {$\displaystyle \widetilde{\Omega}_{6,+}^{(3)}$};
    \end{tikzpicture}
      }}}$%
    }
    \end{equation}
    where the braces collect the connected graphs into $\widetilde\Omega^{(L)}_{6,+}$. Disconnected graphs factorize into products of lower-loop integrands. For labeled loops, the connected integrands are therefore
    \begin{align}
    \widetilde\Omega_{6,+}^{(2)}(a,b)
    ={}&\Omega_{6,+}^{(2)}(a,b)
    -\Omega_{6,+}^{(1)}(a)\Omega_{6,+}^{(1)}(b),\label{eq:connected2}\\
    \widetilde\Omega_{6,+}^{(3)}(a,b,c)
    ={}&\Omega_{6,+}^{(3)}(a,b,c)
    -\sum_{\mathrm{cyc}(a,b,c)}\Omega_{6,+}^{(1)}(a)
    \Omega_{6,+}^{(2)}(b,c)\notag\\
    &+2\,\Omega_{6,+}^{(1)}(a)\Omega_{6,+}^{(1)}(b)
    \Omega_{6,+}^{(1)}(c),\label{eq:labeled}
    \end{align}
    and the same relations hold on the negative branch. The connected integrands therefore give the logarithm of the amplitude at the integrand level,
    \begin{equation}\label{eq:log}
    \log M_\pm=\sum_{L\geq1}g^L\,\frac1{L!}\int_{a_1,\dots,a_L}\widetilde\Omega^{(L)}_{6,\pm},
    \end{equation}
    where $M_\pm=1+\sum_{L\geq1}g^LM_\pm^{(L)}$.
    
    In ABJM theory, only bipartite negative geometries survive~\cite{He:2022cup,He:2023rou}; in particular, the triangle graph in Eq.~\eqref{eq:ng-expansion} cancels, and the connected three-loop integrand has a unique chain topology. At six points, black loops have propagators only to the odd dual points $x_{1,3,5}$ and white loops only to the even ones $x_{2,4,6}$. Summing over the two colorings and the three choices of the middle loop,
    \begin{equation}\label{eq:3loopg}
    \adjustbox{max width=0.97\columnwidth}{%
      $\displaystyle \widetilde\Omega_{6,+}^{(3)}=\vcenter{\hbox{\scalebox{0.8}
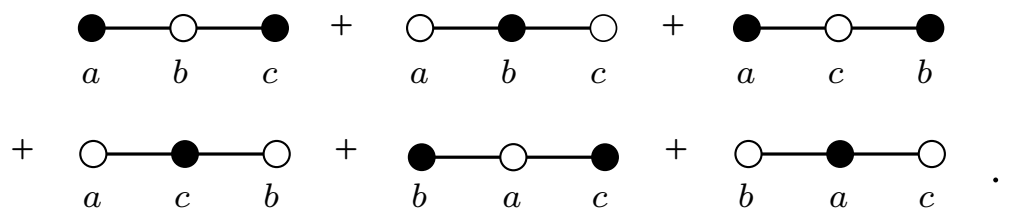
{%
    \begin{tikzpicture}[x=0.75pt,y=0.75pt,yscale=-1,xscale=1]
    % Row 1
    \draw [line width=0.75]    (59.9,119.4) -- (117.4,119.4) ;
    \draw  [fill={rgb, 255:red, 0; green, 0; blue, 0 }  ,fill opacity=1 ] (113.3,119.4) .. controls (113.3,117.14) and (115.14,115.3) .. (117.4,115.3) .. controls (119.66,115.3) and (121.5,117.14) .. (121.5,119.4) .. controls (121.5,121.66) and (119.66,123.5) .. (117.4,123.5) .. controls (115.14,123.5) and (113.3,121.66) .. (113.3,119.4) -- cycle ;
    \draw  [fill={rgb, 255:red, 0; green, 0; blue, 0 }  ,fill opacity=1 ] (55.8,119.4) .. controls (55.8,117.14) and (57.64,115.3) .. (59.9,115.3) .. controls (62.16,115.3) and (64,117.14) .. (64,119.4) .. controls (64,121.66) and (62.16,123.5) .. (59.9,123.5) .. controls (57.64,123.5) and (55.8,121.66) .. (55.8,119.4) -- cycle ;
    \draw  [line width=0.5][fill={rgb, 255:red, 255; green, 255; blue, 255 }  ,fill opacity=1 ] (84.55,119.4) .. controls (84.55,117.14) and (86.39,115.3) .. (88.65,115.3) .. controls (90.91,115.3) and (92.75,117.14) .. (92.75,119.4) .. controls (92.75,121.66) and (90.91,123.5) .. (88.65,123.5) .. controls (86.39,123.5) and (84.55,121.66) .. (84.55,119.4) -- cycle ;
    \draw [line width=0.75]    (162.9,119.4) -- (220.4,119.4) ;
    \draw  [fill={rgb, 255:red, 255; green, 255; blue, 255 }  ,fill opacity=1 ] (216.3,119.4) .. controls (216.3,117.14) and (218.14,115.3) .. (220.4,115.3) .. controls (222.66,115.3) and (224.5,117.14) .. (224.5,119.4) .. controls (224.5,121.66) and (222.66,123.5) .. (220.4,123.5) .. controls (218.14,123.5) and (216.3,121.66) .. (216.3,119.4) -- cycle ;
    \draw  [line width=0.5][fill={rgb, 255:red, 255; green, 255; blue, 255 }  ,fill opacity=1 ] (158.8,119.4) .. controls (158.8,117.14) and (160.64,115.3) .. (162.9,115.3) .. controls (165.16,115.3) and (167,117.14) .. (167,119.4) .. controls (167,121.66) and (165.16,123.5) .. (162.9,123.5) .. controls (160.64,123.5) and (158.8,121.66) .. (158.8,119.4) -- cycle ;
    \draw  [fill={rgb, 255:red, 0; green, 0; blue, 0 }  ,fill opacity=1 ] (187.55,119.4) .. controls (187.55,117.14) and (189.39,115.3) .. (191.65,115.3) .. controls (193.91,115.3) and (195.75,117.14) .. (195.75,119.4) .. controls (195.75,121.66) and (193.91,123.5) .. (191.65,123.5) .. controls (189.39,123.5) and (187.55,121.66) .. (187.55,119.4) -- cycle ;
    \draw [line width=0.75]    (265.4,119.4) -- (322.9,119.4) ;
    \draw  [line width=0.5][fill={rgb, 255:red, 0; green, 0; blue, 0 }  ,fill opacity=1 ] (318.8,119.4) .. controls (318.8,117.14) and (320.64,115.3) .. (322.9,115.3) .. controls (325.16,115.3) and (327,117.14) .. (327,119.4) .. controls (327,121.66) and (325.16,123.5) .. (322.9,123.5) .. controls (320.64,123.5) and (318.8,121.66) .. (318.8,119.4) -- cycle ;
    \draw  [fill={rgb, 255:red, 0; green, 0; blue, 0 }  ,fill opacity=1 ] (261.3,119.4) .. controls (261.3,117.14) and (263.14,115.3) .. (265.4,115.3) .. controls (267.66,115.3) and (269.5,117.14) .. (269.5,119.4) .. controls (269.5,121.66) and (267.66,123.5) .. (265.4,123.5) .. controls (263.14,123.5) and (261.3,121.66) .. (261.3,119.4) -- cycle ;
    \draw  [line width=0.5][fill={rgb, 255:red, 255; green, 255; blue, 255 }  ,fill opacity=1 ] (290.05,119.4) .. controls (290.05,117.14) and (291.89,115.3) .. (294.15,115.3) .. controls (296.41,115.3) and (298.25,117.14) .. (298.25,119.4) .. controls (298.25,121.66) and (296.41,123.5) .. (294.15,123.5) .. controls (291.89,123.5) and (290.05,121.66) .. (290.05,119.4) -- cycle ;
    % Row 2
    \draw [line width=0.75]    (60.4,158.9) -- (117.9,158.9) ;
    \draw  [line width=0.5][fill={rgb, 255:red, 255; green, 255; blue, 255 }  ,fill opacity=1 ] (113.8,158.9) .. controls (113.8,156.64) and (115.64,154.8) .. (117.9,154.8) .. controls (120.16,154.8) and (122,156.64) .. (122,158.9) .. controls (122,161.16) and (120.16,163) .. (117.9,163) .. controls (115.64,163) and (113.8,161.16) .. (113.8,158.9) -- cycle ;
    \draw  [line width=0.5][fill={rgb, 255:red, 255; green, 255; blue, 255 }  ,fill opacity=1 ] (56.3,158.9) .. controls (56.3,156.64) and (58.14,154.8) .. (60.4,154.8) .. controls (62.66,154.8) and (64.5,156.64) .. (64.5,158.9) .. controls (64.5,161.16) and (62.66,163) .. (60.4,163) .. controls (58.14,163) and (56.3,161.16) .. (56.3,158.9) -- cycle ;
    \draw  [fill={rgb, 255:red, 0; green, 0; blue, 0 }  ,fill opacity=1 ] (85.05,158.9) .. controls (85.05,156.64) and (86.89,154.8) .. (89.15,154.8) .. controls (91.41,154.8) and (93.25,156.64) .. (93.25,158.9) .. controls (93.25,161.16) and (91.41,163) .. (89.15,163) .. controls (86.89,163) and (85.05,161.16) .. (85.05,158.9) -- cycle ;
    \draw [line width=0.75]    (163.4,159.9) -- (220.9,159.9) ;
    \draw  [fill={rgb, 255:red, 0; green, 0; blue, 0 }  ,fill opacity=1 ] (216.8,159.9) .. controls (216.8,157.64) and (218.64,155.8) .. (220.9,155.8) .. controls (223.16,155.8) and (225,157.64) .. (225,159.9) .. controls (225,162.16) and (223.16,164) .. (220.9,164) .. controls (218.64,164) and (216.8,162.16) .. (216.8,159.9) -- cycle ;
    \draw  [fill={rgb, 255:red, 0; green, 0; blue, 0 }  ,fill opacity=1 ] (159.3,159.9) .. controls (159.3,157.64) and (161.14,155.8) .. (163.4,155.8) .. controls (165.66,155.8) and (167.5,157.64) .. (167.5,159.9) .. controls (167.5,162.16) and (165.66,164) .. (163.4,164) .. controls (161.14,164) and (159.3,162.16) .. (159.3,159.9) -- cycle ;
    \draw  [line width=0.5][fill={rgb, 255:red, 255; green, 255; blue, 255 }  ,fill opacity=1 ] (188.05,159.9) .. controls (188.05,157.64) and (189.89,155.8) .. (192.15,155.8) .. controls (194.41,155.8) and (196.25,157.64) .. (196.25,159.9) .. controls (196.25,162.16) and (194.41,164) .. (192.15,164) .. controls (189.89,164) and (188.05,162.16) .. (188.05,159.9) -- cycle ;
    \draw [line width=0.75]    (265.9,158.9) -- (323.4,158.9) ;
    \draw  [line width=0.5][fill={rgb, 255:red, 255; green, 255; blue, 255 }  ,fill opacity=1 ] (319.3,158.9) .. controls (319.3,156.64) and (321.14,154.8) .. (323.4,154.8) .. controls (325.66,154.8) and (327.5,156.64) .. (327.5,158.9) .. controls (327.5,161.16) and (325.66,163) .. (323.4,163) .. controls (321.14,163) and (319.3,161.16) .. (319.3,158.9) -- cycle ;
    \draw  [line width=0.5][fill={rgb, 255:red, 255; green, 255; blue, 255 }  ,fill opacity=1 ] (261.8,158.9) .. controls (261.8,156.64) and (263.64,154.8) .. (265.9,154.8) .. controls (268.16,154.8) and (270,156.64) .. (270,158.9) .. controls (270,161.16) and (268.16,163) .. (265.9,163) .. controls (263.64,163) and (261.8,161.16) .. (261.8,158.9) -- cycle ;
    \draw  [fill={rgb, 255:red, 0; green, 0; blue, 0 }  ,fill opacity=1 ] (290.55,158.9) .. controls (290.55,156.64) and (292.39,154.8) .. (294.65,154.8) .. controls (296.91,154.8) and (298.75,156.64) .. (298.75,158.9) .. controls (298.75,161.16) and (296.91,163) .. (294.65,163) .. controls (292.39,163) and (290.55,161.16) .. (290.55,158.9) -- cycle ;
    % Loop labels and signs
    \draw (132.98,114) node [anchor=north west][inner sep=0.75pt]  [font=\normalsize] [align=left] {$\displaystyle +$};
    \draw (55.38,136.5) node [anchor=base west][inner sep=0.75pt]  [font=\footnotesize] [align=left] {$\displaystyle a$};
    \draw (83.98,136.5) node [anchor=base west][inner sep=0.75pt]  [font=\footnotesize] [align=left] {$\displaystyle b$};
    \draw (111.98,136.5) node [anchor=base west][inner sep=0.75pt]  [font=\footnotesize] [align=left] {$\displaystyle c$};
    \draw (236.98,114) node [anchor=north west][inner sep=0.75pt]  [font=\normalsize] [align=left] {$\displaystyle +$};
    \draw (317.48,136.5) node [anchor=base west][inner sep=0.75pt]  [font=\footnotesize] [align=left] {$\displaystyle b$};
    \draw (289.48,136.5) node [anchor=base west][inner sep=0.75pt]  [font=\footnotesize] [align=left] {$\displaystyle c$};
    \draw (260.88,136.5) node [anchor=base west][inner sep=0.75pt]  [font=\footnotesize] [align=left] {$\displaystyle a$};
    \draw (134.48,153) node [anchor=north west][inner sep=0.75pt]  [font=\normalsize] [align=left] {$\displaystyle +$};
    \draw (237.98,153) node [anchor=north west][inner sep=0.75pt]  [font=\normalsize] [align=left] {$\displaystyle +$};
    \draw (215.48,175.6) node [anchor=base west][inner sep=0.75pt]  [font=\footnotesize] [align=left] {$\displaystyle c$};
    \draw (187.48,175.6) node [anchor=base west][inner sep=0.75pt]  [font=\footnotesize] [align=left] {$\displaystyle a$};
    \draw (158.88,175.6) node [anchor=base west][inner sep=0.75pt]  [font=\footnotesize] [align=left] {$\displaystyle b$};
    \draw (32.98,153) node [anchor=north west][inner sep=0.75pt]  [font=\normalsize] [align=left] {$\displaystyle +$};
    \draw (158.38,136.5) node [anchor=base west][inner sep=0.75pt]  [font=\footnotesize] [align=left] {$\displaystyle a$};
    \draw (186.98,136.5) node [anchor=base west][inner sep=0.75pt]  [font=\footnotesize] [align=left] {$\displaystyle b$};
    \draw (214.98,136.5) node [anchor=base west][inner sep=0.75pt]  [font=\footnotesize] [align=left] {$\displaystyle c$};
    \draw (261.38,175.6) node [anchor=base west][inner sep=0.75pt]  [font=\footnotesize] [align=left] {$\displaystyle b$};
    \draw (289.98,175.6) node [anchor=base west][inner sep=0.75pt]  [font=\footnotesize] [align=left] {$\displaystyle a$};
    \draw (317.98,175.6) node [anchor=base west][inner sep=0.75pt]  [font=\footnotesize] [align=left] {$\displaystyle c$};
    \draw (55.88,175.6) node [anchor=base west][inner sep=0.75pt]  [font=\footnotesize] [align=left] {$\displaystyle a$};
    \draw (84.48,175.6) node [anchor=base west][inner sep=0.75pt]  [font=\footnotesize] [align=left] {$\displaystyle c$};
    \draw (112.48,175.6) node [anchor=base west][inner sep=0.75pt]  [font=\footnotesize] [align=left] {$\displaystyle b$};
    \draw (340,165) node [anchor=north west][inner sep=0.75pt]  [font=\large] [align=left] {$\displaystyle .$};
    \end{tikzpicture}
      }}}$%
    }
    \end{equation}
    where each graph stands for its propagators, including a mutual propagator for each edge, multiplied by a numerator. The first graph has the propagators
    \begin{equation}\label{eq:chain-denom}
    D(a,b,c)=(a\cdot b)(b\cdot c)\prod_{i=1,3,5}(a\cdot i)(c\cdot i)\prod_{j=2,4,6}(b\cdot j),
    \end{equation}
    and the others follow by relabeling.
    
    \emph{Integrand from unitarity.}---We construct $\widetilde\Omega^{(3)}_{6,+}$ from a dual-conformal ansatz with the propagators~\eqref{eq:chain-denom}. Dual conformal invariance fixes the numerator weights to $(1,2,1)$ in $(y_a,y_b,y_c)$. Besides the $(i\cdot j)$, the numerators may contain embedding-space $\epsilon$ tensors and the spinor products $\langle 12\rangle\langle 34\rangle\langle 56\rangle$ and $\langle 23\rangle\langle 45\rangle\langle 61\rangle$, which carry unit weight in each odd and each even external point, respectively. We fix the coefficients with two-particle cuts. On the cut $(a\cdot 1)=(a\cdot 3)=0$, with solution $x_{a^\star}$, the integrand factorizes into products of lower-loop four- and six-point integrands~\cite{He:2023rou},
    \begin{multline}\label{eq:cut}
    \Omega^{(L)}_{6,+}\big|_{\mathrm{cut}}
    =\Omega^{(1)}_{6,+}\big|_{\mathrm{cut}}
    \sum_{k=0}^{L-1}\Omega^{(k)}_{4}(x_1,x_2,x_3,x_{a^\star})\\
    \times\Omega^{(L-1-k)}_{6,+}(x_1,x_{a^\star},x_3,x_4,x_5,x_6),
    \end{multline}
    where the remaining $L-1$ loop labels are distributed in all possible ways between the four- and six-point integrands on the two sides of the cut, and the one-loop cut supplies the tree-level factor, including the sum over internal states. Taking the same cut of Eq.~\eqref{eq:labeled}, with the disconnected products evaluated at the original external points $x_1,\dots,x_6$, gives the cut that the ansatz for $\widetilde\Omega^{(3)}_{6,+}$ must reproduce.

    We split the resulting integrand into parity-even and parity-odd parts, which contain an even and an odd number of $\epsilon$ tensors, respectively. This grading is distinct from the parity transformation that exchanges the two branches. We write
    \begin{equation}\label{eq:full-integrand}
    \widetilde\Omega^{(3)}_{6,+}=\widetilde\Omega^{(3),\mathrm{even}}_{6,+}+\widetilde\Omega^{(3),\mathrm{odd}}_{6,+},
    \end{equation}
    and generate each part from an integrand with black endpoint loops $a$, $c$ and a white middle loop $b$,
    \begin{multline}\label{eq:parity-functions}
    \widetilde\Omega^{(3),\mathrm{even}}_{6,+}=-\sum_{\mathrm{perms}(a,b,c)}\bigl[I^{\mathrm{even}}+(i\mapsto i+2k)\\
    +(1\leftrightarrow2,\,3\leftrightarrow6,\,4\leftrightarrow5)\bigr],
    \end{multline}
    and likewise $\widetilde\Omega^{(3),\mathrm{odd}}_{6,+}$ from $I^{\mathrm{odd}}$. The sum runs over the six permutations of the loop labels, $(i\mapsto i+2k)$ denotes the two additional cyclic images ($k=1,2$), and the reflection acts on the full cyclic sum, exchanging $\langle 12\rangle\langle 34\rangle\langle 56\rangle$ and $\langle 23\rangle\langle 45\rangle\langle 61\rangle$. Explicit expressions for $I^{\mathrm{even}}$ and $I^{\mathrm{odd}}$, each a sum of ten terms, are given in Appendix~\ref{app:simple-finite-integrands} of the Supplemental Material (SM).

\section{Complete integration of negative geometries}\label{sec:integration}

Negative geometries are usually integrated over all but one loop. This yields an infrared-finite function of the remaining loop variable, from which infrared quantities such as the cusp anomalous dimension can be extracted~\cite{Arkani-Hamed:2021iya,He:2023exb,Henn:2023pkc,Lagares:2024epo,Li:2024lbw}. The logarithm of the amplitude, Eq.~\eqref{eq:log}, instead requires integrating over all $L$ loops, which in general produces infrared divergences. In ABJM theory, however, we find that the fully integrated negative geometry remains finite at three loops. This is consistent with two earlier results: the cusp anomalous dimension, which controls the leading infrared divergence at each loop order, contains only even powers of $N/k$~\cite{Gromov:2008qe,Gromov:2014eha}, and at four points the logarithm of the amplitude has no three-loop divergence~\cite{Bianchi:2014iia}. Below we derive this result by first integrating out the two endpoint loops of the chain and then the middle loop.

\begin{figure}[t]
\centering
\begin{tikzpicture}[x=.90cm,y=.90cm,>=Latex]
\tikzset{blackv/.style={circle,fill=black,inner sep=2.3pt},whitev/.style={circle,draw,fill=white,inner sep=2.3pt}}
\node[blackv,label=above:$a$,label={[font=\scriptsize]below:$1,3,5$}] (a) at (0,0) {};
\node[whitev,label=above:$b$,label={[font=\scriptsize]below:$2,4,6$}] (b) at (1.2,0) {};
\node[blackv,label=above:$c$,label={[font=\scriptsize]below:$1,3,5$}] (c) at (2.4,0) {};
\draw (a)--(b)--(c);
\draw[->] (3,0)--node[above,font=\small]{$\int_{a,c}$}(4.3,0);
\node[blackv,label=above:$b$,label={[font=\scriptsize]below:$1,3,5$}] at (4.9,0) {};
\node at (5.65,0) {$+$};
\node[whitev,label=above:$b$,label={[font=\scriptsize]below:$2,4,6$}] at (6.4,0) {};
\end{tikzpicture}
\caption{Integrating out the endpoint loops $a$ and $c$ reduces the three-loop chain to $-2\pi^2$ times the one-loop integrand $\Omega^{(1)}_{6,+}(b)$. Only its scalar triangles are drawn, since the $\epsilon$-tensor boxes integrate to zero. Labels below each vertex specify its external poles; an edge supplies a mutual propagator.}
\label{fig:ladder}
\end{figure}
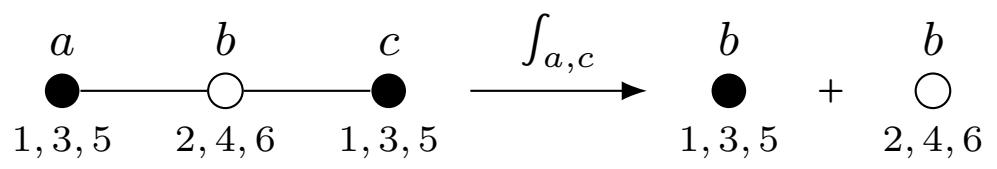

\emph{Endpoint integration.}---The chain topology suggests integrating out the two endpoint loops first: $a$ and $c$ are attached to the rest of the graph only through the propagators $(a\cdot b)$ and $(b\cdot c)$, and the numerators of $I^{\mathrm{even}}$ and $I^{\mathrm{odd}}$ are linear in $y_a$ and in $y_c$. For fixed $b$, each endpoint integral is therefore a one-loop tensor box or, when the endpoint has only three propagators, a scalar triangle. Since every term of $I^{\mathrm{even}}$ and $I^{\mathrm{odd}}$ is locally finite (see Appendix~\ref{app:simple-finite-integrands} of the SM), the connected integral converges absolutely for generic Euclidean kinematics, so it is finite and the order of integration is immaterial. For a black endpoint, we find in the Euclidean region
\begin{multline}\label{eq:endpoint}
\int_a\frac{y_a^\mu}{(a\cdot 1)(a\cdot 3)(a\cdot 5)(a\cdot b)}\\
=\frac{\pi}{2}\,\frac{K^\mu}{\sqrt{(1\cdot 3)(1\cdot 5)(3\cdot 5)}\,\rho_1\rho_3\rho_5\,(\rho_1+\rho_3+\rho_5)},
\end{multline}
where $\rho_1=\sqrt{(3\cdot 5)(b\cdot 1)}$, $\rho_3=\sqrt{(1\cdot 5)(b\cdot 3)}$, $\rho_5=\sqrt{(1\cdot 3)(b\cdot 5)}$, and
\begin{multline}\label{eq:K}
K^\mu=(3\cdot 5)\rho_3\rho_5\,y_1^\mu+(1\cdot 5)\rho_1\rho_5\,y_3^\mu+(1\cdot 3)\rho_1\rho_3\,y_5^\mu\\
+(1\cdot 3)(1\cdot 5)(3\cdot 5)\,y_b^\mu;
\end{multline}
white endpoints follow by reflection. Since the tensor box involves no $\epsilon$ tensor, it lies in the span of $y_1$, $y_3$, $y_5$, and $y_b$. Because $y_a\cdot y_i=(a\cdot i)$, contracting it with each of these vectors cancels one propagator and leaves a scalar triangle, $\int_a1/[(a\cdot i)(a\cdot j)(a\cdot k)]=\pi/[2\sqrt{(i\cdot j)(j\cdot k)(k\cdot i)}]$, and these four contractions fix the coefficients in Eq.~\eqref{eq:K}. Equation~\eqref{eq:endpoint} thus effectively replaces $y_a$ and $y_c$ by the corresponding vectors $K^\mu$, up to scalar prefactors, and a term in which $a$ enters only through $\epsilon(1,3,5,a,b)$ integrates to zero, since this $\epsilon$ tensor vanishes when $y_a$ is replaced by $y_1$, $y_3$, $y_5$, or $y_b$. Individual terms then depend on $b$ through square roots such as the $\rho_i$, which cancel after summing over the cyclic images and the exchange of the endpoints. In both parts, the two colorings contribute equally, and we write $|_b$ for the terms in which $b$ is the middle loop.

\emph{Parity-even part.}---The last term of $I^{\mathrm{even}}$ drops out, and the vanishing of the six-point Gram determinant and the identities relating the spinor products to the $(i\cdot j)$ reduce the remaining terms to one-loop triangles,
\begin{multline}\label{eq:collapse}
\int_{a,c}\widetilde\Omega^{(3),\mathrm{even}}_{6,+}\Big|_b\\
=-2\pi^2\left[\frac{\langle 12\rangle\langle 34\rangle\langle 56\rangle}{(b\cdot 1)(b\cdot 3)(b\cdot 5)}+\frac{\langle 23\rangle\langle 45\rangle\langle 61\rangle}{(b\cdot 2)(b\cdot 4)(b\cdot 6)}\right].
\end{multline}
Up to the factor $-2\pi^2$, the right-hand side is precisely the triangle part of the one-loop integrand $\Omega^{(1)}_{6,+}(b)$, as illustrated in Fig.~\ref{fig:ladder}. Integrated over $b$, these triangles give the one-loop amplitude
\begin{multline}\label{eq:one-loop}
M_\pm^{(1)}=\mp\frac{i\pi}{2}\bigl[\operatorname{sgn}_c\langle12\rangle\operatorname{sgn}_c\langle34\rangle\operatorname{sgn}_c\langle56\rangle\\
+\operatorname{sgn}_c\langle23\rangle\operatorname{sgn}_c\langle45\rangle\operatorname{sgn}_c\langle61\rangle\bigr],
\end{multline}
with the sign function $\operatorname{sgn}_c$ of Ref.~\cite{Caron-Huot:2012sos}; see Appendix~\ref{app:lower-loops} of the SM.

\emph{Parity-odd part.}---The first two terms of $I^{\mathrm{odd}}$ drop out for the same reason, and the remaining terms reduce to $\epsilon$-tensor boxes,
\begin{multline}\label{eq:collapse-odd}
\int_{a,c}\widetilde\Omega^{(3),\mathrm{odd}}_{6,+}\Big|_b
=-\sqrt2\,\pi^2\biggl[\frac{\epsilon(b,3,4,5,1)}{(b\cdot 3)(b\cdot 4)(b\cdot 5)(b\cdot 1)}\\
+\frac{\epsilon(b,1,2,3,4)}{(b\cdot 1)(b\cdot 2)(b\cdot 3)(b\cdot 4)}
-\frac{\epsilon(b,4,5,6,1)}{(b\cdot 4)(b\cdot 5)(b\cdot 6)(b\cdot 1)}\\
-\frac{\epsilon(b,6,1,2,4)}{(b\cdot 6)(b\cdot 1)(b\cdot 2)(b\cdot 4)}\biggr].
\end{multline}
Up to the same factor $-2\pi^2$, the right-hand side is precisely the box part of $\Omega^{(1)}_{6,+}(b)$. Together with Eq.~\eqref{eq:collapse}, the three-loop chain therefore collapses to the full one-loop integrand,
\begin{equation}\label{eq:collapse-full}
\int_{a,c}\widetilde\Omega^{(3)}_{6,+}\Big|_b=-2\pi^2\,\Omega^{(1)}_{6,+}(b),
\end{equation}
 As at one loop, the $\epsilon$-tensor boxes integrate to zero in three dimensions, so the parity-odd part does not contribute,
\begin{equation}\label{eq:odd-zero}
\int_{a,b,c}\widetilde\Omega^{(3),\mathrm{odd}}_{6,+}=0.
\end{equation}

\emph{Combined result.}---The negative branch follows by flipping the signs of both spinor products. Every term of the parity-even part contains exactly one of them, and every term of the parity-odd part contains neither or both, so that $\widetilde\Omega^{(3)}_{6,-}=-\widetilde\Omega^{(3),\mathrm{even}}_{6,+}+\widetilde\Omega^{(3),\mathrm{odd}}_{6,+}$. The same flip reverses the triangles of the one-loop integrand but not its boxes, so Eq.~\eqref{eq:collapse-full} holds on both branches, with $\Omega^{(1)}_{6,\pm}(b)$ on the right-hand side. Integrating over $b$ and summing over the three choices of the middle loop, we obtain on both branches
\begin{multline}\label{eq:connected-result}
\frac1{3!}\int_{a,b,c}\widetilde\Omega^{(3)}_{6,\pm}
=-\frac{3\times2\pi^2}{3!}\,M_\pm^{(1)}=-\pi^2M_\pm^{(1)}\\
=\pm\frac{i\pi^3}{2}\bigl[\operatorname{sgn}_c\langle12\rangle\operatorname{sgn}_c\langle34\rangle\operatorname{sgn}_c\langle56\rangle\\
+\operatorname{sgn}_c\langle23\rangle\operatorname{sgn}_c\langle45\rangle\operatorname{sgn}_c\langle61\rangle\bigr].
\end{multline}
The connected integral is thus finite and depends on the external kinematics only through the signs of the spinor brackets.

\section{Three-loop amplitude and exponentiation}

By the negative-geometry expansion~\eqref{eq:log}, the connected integral~\eqref{eq:connected-result} is the three-loop contribution to $\log M_\pm$, which establishes Eq.~\eqref{eq:prl-intro-exponentiation}. Expanding $M_\pm=\exp(\log M_\pm)$ to three loops then gives $M^{(3)}_\pm=M_\pm^{(1)}M_\pm^{(2)}-\frac13(M_\pm^{(1)})^3-\pi^2M_\pm^{(1)}$. In physical kinematics each $\operatorname{sgn}_c$ equals $\pm1$, so $M^{(1)}_\pm$ takes only the values $0$ and $\pm i\pi$, and $(M^{(1)}_\pm)^3=-\pi^2M^{(1)}_\pm$. Hence
\begin{equation}\label{eq:M3}
M_\pm^{(3)}=M_\pm^{(1)}\Bigl[M_\pm^{(2)}-\frac{2\pi^2}{3}\Bigr].
\end{equation}

In mass regularization, the two-loop result of Ref.~\cite{Caron-Huot:2012sos} reads
\begin{equation}\label{eq:M2}
M_\pm^{(2)}=\frac12\bigl[\mathrm{BDS}_6+R_6\bigr]\mp\frac i2F,
\end{equation}
where the BDS ansatz $\mathrm{BDS}_6$~\cite{Bern:2005iz} contains all infrared divergences, $R_6$ is the dual-conformally invariant remainder function (both given in Appendix~\ref{app:lower-loops} of the SM), and
\begin{multline}\label{eq:F}
F=\operatorname{sgn}_c\langle12\rangle\operatorname{sgn}_c\langle45\rangle
\operatorname{sgn}_c\bigl[i(\langle34\rangle\langle46\rangle+\langle35\rangle\langle56\rangle)\bigr]\\
\times\log\frac{u_2}{u_3}\arccos\sqrt{u_1}+\text{cyclic}\times2.
\end{multline}
Dressing the two branches with the leading singularities $\mathrm{LS}_{6,\pm}$, we obtain the three-loop amplitude
\begin{multline}\label{eq:final-amplitude}
\mathcal A_6^{(3)}=-\frac\pi4\bigl[\operatorname{sgn}_c\langle12\rangle\operatorname{sgn}_c\langle34\rangle\operatorname{sgn}_c\langle56\rangle\\
+\operatorname{sgn}_c\langle23\rangle\operatorname{sgn}_c\langle45\rangle\operatorname{sgn}_c\langle61\rangle\bigr]\\
\times\Bigl\{\mathcal A_{6,\mathrm{shifted}}^{\mathrm{tree}}\Bigl[\mathrm{BDS}_6+R_6-\frac{4\pi^2}{3}\Bigr]+\mathcal A_6^{\mathrm{tree}}\,F\Bigr\}.
\end{multline}

The first term in Eq.~\eqref{eq:final-amplitude} equals $\frac12\mathcal A_6^{(1)}[\mathrm{BDS}_6+R_6-4\pi^2/3]$, since the one-loop amplitude is $-\frac\pi2\mathcal A_{6,\mathrm{shifted}}^{\mathrm{tree}}$ times the bracket in Eq.~\eqref{eq:one-loop}. Beyond the iteration of lower-loop amplitudes, the only genuinely three-loop information is the constant $-\pi^2$ in Eq.~\eqref{eq:prl-intro-exponentiation}. The two-loop functions $R_6$ and $F$ thus reappear with the one-loop sign factor, but with their Yangian invariants exchanged: $R_6$ now multiplies the shifted tree and $F$ the tree amplitude.

\emph{Implications of the three-loop result.}---  Our result confirms the expectation of Ref.~\cite{Caron-Huot:2012sos} that the three-loop amplitude has uniform transcendental weight three (counting $\pi$ as weight one) and that its infrared divergences multiply only the one-loop amplitude. Since the infrared divergences of $\log M_\pm$ are controlled by the cusp and collinear anomalous dimensions, the finiteness of $\log M_\pm$ at three loops implies that both have vanishing three-loop coefficients. For the cusp anomalous dimension this agrees with the all-loop prediction from integrability~\cite{Gromov:2008qe,Gromov:2014eha}, while for the collinear anomalous dimension it extends the four-point evidence of Ref.~\cite{Bianchi:2014iia} to six points.

%\ky{KY: added some discussion,feel free to remove.} 
Moreover, our results suggest an iterative relation between the odd- and even-loop contributions. Through three loops, the perturbative amplitude can be organized into a concise form,
\begin{equation}
\mathcal A_6(g)=\bigl[1+g\,f(g^2)\,\mathbb M_1\bigr]\,\mathcal A_6^{\mathrm{even}}(g^2),
\end{equation}
where $\mathbb M_1$ is an operator satisfying
\begin{align}
\mathbb M_1\,\mathcal A_6^{\mathrm{tree}}&=-s_{\mathrm{hex}}\,\pi\,\mathcal A_{6,\mathrm{shifted}}^{\mathrm{tree}},\notag\\
\mathbb M_1\,\mathcal A_{6,\mathrm{shifted}}^{\mathrm{tree}}&=s_{\mathrm{hex}}\,\pi\,\mathcal A_6^{\mathrm{tree}}.
\end{align}
Here $s_{\mathrm{hex}}\in\{0,\pm1\}$ is defined by $M_\pm^{(1)}=\mp i\pi s_{\mathrm{hex}}$, $\mathcal A_6(g)$ is the amplitude truncated at $\mathcal O(g^3)$, and $\mathcal A_6^{\mathrm{even}}(g^2)$ is the sum of its even-loop terms. Our computation determines the leading correction to the function $f$,
\begin{equation}
f(g^2)=1-\frac{2\pi^2}{3}\,g^2+\mathcal O(g^4).
\end{equation}
The absence of kinematic dependence in $f$ allows this correction to be absorbed into an effective coupling $g\to\widehat g=g\,f(g^2)$. As a consequence of the three-loop negative geometry, the odd-loop amplitudes inherit both their kinematic dependence and infrared divergences from the even-loop sector, through $\mathbb M_1$ supplemented by a kinematic-independent function.
It remains an interesting open question whether a finite, constant function $f$ continues to organize the amplitudes (or a subset of diagrams) at 
higher loop orders.  
%If it does, even-loop sector determines both the IR divergences and kinematic dependence of the full amplitudes
%the kinematic dependence in the odd-loop amplitude will be generated  from the
%even-loop sector through $\mathbb M_1$ supplemented by a
%kinematic-independent function, whose value can be determined by matching higher-loop IR divergences onto lower loop data.\cite{Grozin:2014hna}. 

\section{Discussion}

We have computed the three-loop six-point amplitude of ABJM theory by integrating the connected negative geometry over all three loops. The result is strikingly simple: even though six points is the first multiplicity with nontrivial cross-ratios, the three-loop contribution to $\log M_\pm$ is a pure number times the one-loop amplitude. It would be interesting to understand whether the odd-loop part of $\log M_\pm$ remains proportional to the one-loop amplitude at higher loops, for instance through a similar collapse of the connected integrand to $\Omega^{(1)}_{6,\pm}$.

The function space is equally simple. Although the three-loop amplitude has uniform transcendental weight three, it contains no genuinely weight-three polylogarithms: every term in Eq.~\eqref{eq:final-amplitude} is $\pi$ times a weight-two function already present at two loops. New functions can therefore first appear at four loops. Given this simplicity, the relevant function space may be considerably smaller than that of generic weight-four polylogarithms, making a bootstrap of the four-loop six-point amplitude promising.

At four loops, negative geometries offer a complementary handle: the connected bipartite topologies are the chain, the star, and the box~\cite{He:2023rou}. The first two are trees, and Eq.~\eqref{eq:endpoint} applies to each of their leaves, suggesting a leaf-by-leaf integration similar to that of the ladder-type negative geometries at four points~\cite{He:2023exb,Li:2024lbw}. At this even loop order, however, the connected geometries contribute to the cusp anomalous dimension, so their full integration is infrared divergent. Integrating out all but one loop is nevertheless expected to yield infrared-finite six-point functions of the remaining loop variable, which generalize the infrared-finite four-point observables of Refs.~\cite{He:2023exb,Henn:2023pkc} and will be studied in future work.

Our approach should also extend to higher multiplicity: by bipartiteness, the connected three-loop geometry remains a chain, and its endpoint loops can again be integrated out first, since one-loop integrals in three dimensions reduce to triangles. Moreover, the cusp and collinear anomalous dimensions are universal, so their vanishing at three loops implies that the fully integrated three-loop negative geometry is infrared finite at any multiplicity. More generally, since the cusp anomalous dimension contains only even powers of $N/k$~\cite{Gromov:2008qe,Gromov:2014eha}, we conjecture that the fully integrated connected negative geometries are infrared finite at every odd loop order, which amounts to the collinear anomalous dimension being even in $N/k$ as well.

\begin{acknowledgments}
We thank Yu-tin Huang and Chi Zhang for helpful discussions.  K. Yan and H. Y. Fang are supported by National Natural Science Foundation of China under Grant No. 12357077. C.-K. Kuo is funded by the European Union (ERC, 101118787). Views and opinions expressed are, however, those of the author(s) only and do not necessarily reflect those of the European Union or the European Research Council Executive Agency. Neither the European Union nor the granting authority can be held responsible for them.
\end{acknowledgments}

\bibliographystyle{apsrev4-1}
\bibliography{bib}% Produces the bibliography via BibTeX.

\newpage

\appendix

\widetext
\raggedbottom
\begin{center}
\textbf{\Large Supplemental Material}
\end{center}

\section{Lower-loop amplitudes}\label{app:lower-loops}

The mass regulator of Ref.~\cite{Caron-Huot:2012sos} shifts the external embedding vectors as
\begin{equation}\label{eq:mass-regulator}
\widehat y_i=y_i+\mu_{\mathrm{IR}}^2y_I,\qquad
(a\cdot \widehat i)=x_{ai}^2+\mu_{\mathrm{IR}}^2,\qquad
(\widehat i\cdot \widehat j)=x_{ij}^2+2\mu_{\mathrm{IR}}^2,\qquad
(a\cdot b)=x_{ab}^2,
\end{equation}
with $y_I=(0,0,0,0,1)$.

In terms of the one-loop box and triangle integrals
\begin{gather}
I_{\mathrm{box}}(i,j,k,l)\equiv\int_a\frac{\epsilon(a,i,j,k,l)}{(a\cdot i)(a\cdot j)(a\cdot k)(a\cdot l)},\notag\\
I_{\mathrm{tri}}(1,3,5)\equiv\int_a\frac{\langle 12\rangle\langle 34\rangle\langle 56\rangle}{(a\cdot 1)(a\cdot 3)(a\cdot 5)},\qquad
I_{\mathrm{tri}}(2,4,6)\equiv\int_a\frac{\langle 23\rangle\langle 45\rangle\langle 61\rangle}{(a\cdot 2)(a\cdot 4)(a\cdot 6)},
\end{gather}
the one-loop correction on the positive branch reads
\begin{equation}\label{eq:one-loop-input}
M_+^{(1)}=\frac1{\sqrt2}\Bigl[I_{\mathrm{box}}(3,4,5,1)+I_{\mathrm{box}}(1,2,3,4)-I_{\mathrm{box}}(4,5,6,1)-I_{\mathrm{box}}(6,1,2,4)\Bigr]
+I_{\mathrm{tri}}(1,3,5)+I_{\mathrm{tri}}(2,4,6).
\end{equation}
The box integrals vanish in three dimensions, and the triangle integrals give Eq.~\eqref{eq:one-loop}. Restoring the leading singularities, the one-loop amplitude is $\mathcal A_6^{(1)}=-\frac\pi2\mathcal A_{6,\mathrm{shifted}}^{\mathrm{tree}}[\operatorname{sgn}_c\langle 12\rangle\operatorname{sgn}_c\langle 34\rangle\operatorname{sgn}_c\langle 56\rangle
+\operatorname{sgn}_c\langle 23\rangle\operatorname{sgn}_c\langle 45\rangle\operatorname{sgn}_c\langle 61\rangle]$.

At two loops, the BDS ansatz and the remainder function read
\begin{align}\label{eq:BDS}
\mathrm{BDS}_6={}&\frac12\sum_{i=1}^{6}\left[
\log^2\!\frac{x_{i,i+3}^2}{x_{i+1,i+3}^2}
-\log^2\!\frac{4\mu_{\mathrm{IR}}^2x_{i,i+3}^2}{x_{i,i+2}^2x_{i+1,i+3}^2}\right]
+\sum_{i=1}^{3}\left[-\operatorname{Li}_2(1-u_i)+\log u_i\log\!\frac{x_{i-1,i+2}^2}{4\mu_{\mathrm{IR}}^2}\right]+\pi^2,\\
R_6={}&-2\pi^2+\sum_{i=1}^{3}\left[\operatorname{Li}_2(1-u_i)+\frac12\log u_i\log u_{i+1}+\bigl(\arccos\sqrt{u_i}\bigr)^2\right],\label{eq:R6}
\end{align}
with $u_{i+3}\equiv u_i$, and the full two-loop amplitude is
\begin{equation}\label{eq:two-amplitude}
\mathcal A_6^{(2)}=\frac{\mathcal A_6^{\mathrm{tree}}}{2}\bigl[\mathrm{BDS}_6+R_6\bigr]-\frac{\mathcal A_{6,\mathrm{shifted}}^{\mathrm{tree}}}{2}F,
\end{equation}
with $F$ given in Eq.~\eqref{eq:F}. Projecting onto the two leading singularities with $\mathrm{LS}_{6,+}-\mathrm{LS}_{6,-}=-i\mathcal A_{6,\mathrm{shifted}}^{\mathrm{tree}}$ gives Eq.~\eqref{eq:M2}.

\section{Simple finite integrals}
\label{app:simple-finite-integrands}
\subsection{explicit even and odd integrands}

In this appendix, we give the integrands $I^{\rm even}$ and $I^{\rm odd}$ that generate the connected three-loop integrand through Eq.~\eqref{eq:parity-functions}. Both have black endpoint loops $a$, $c$ and a white middle loop $b$. The parity-even integrand $I^{\rm even}$ consists of ten locally finite terms spanning the top topology and eight sub-topologies. Their explicit expressions are grouped by topology below.

%\begin{widetext}
%\begingroup
%\setlength{\jot}{3pt}
\begin{align}
I^{\rm even}  
& ={}
\frac56 \Se \, d_{13}d_{35}d_{15}\,
\BGraph{1,3,5}{2,4,6}{1,3,5}{0}{0}
%\label{line0ae}
 \nonumber
\\
&
+
\frac{\So}2\,
\BGraph{1,3,5}{2,4,6}{1,3,5}{0}{1}
\left[
\frac{\epsilon_{23456}\epsilon_{1234c}\epsilon_{4561b}}
     {\epsilon_{12345}d_{46}} + \frac{\epsilon_{12346}\epsilon_{2345c}\epsilon_{5612b}}
     {\epsilon_{12345}d_{26}} 
\right]
\nonumber
\\
&+
\frac{\Se}{4d_{26}d_{46}d_{24}(u_2^{-1}-1)}\,
\BGraph{1,5}{2,4}{1,5}{1}{1}
\epsilon_{1245b}\epsilon_{12456}
%\label{line1e}
%\\
%&
\quad +
\frac{\Se}{4d_{26}d_{46}}\,
\BGraph{1,5}{2,4,6}{1,5}{1}{1}
\epsilon_{4561b}\epsilon_{5612b}
\nonumber  
\\
&+
\frac{\Se }{2d_{24}d_{46}}\,
\BGraph{1,3}{2,4,6}{1,5}{1}{1}
\epsilon_{1234b}\epsilon_{4561b}
%\label{line3e}
%\\
\quad +
\frac{\So}{2d_{13}d_{25}}\,
\BGraph{1,3,5}{2}{1,3,5}{1}{1}
\left[
\epsilon_{1352a}\epsilon_{1352c}
+2d_{13}d_{35}d_{15}a_2c_2
\right]
%\label{line4e}
\nonumber
%\\
%&+
%\frac{\So}{d_{35}d_{46}}\,
%\BGraph{1,3,5}{4,6}{1,5}{1}{1}
%\epsilon_{3456a}\epsilon_{4561b}
%%\label{line5e}
%%\\
%\quad +
%\frac{\So}{d_{13}d_{26}}\,
%\BGraph{1,3,5}{2,6}{1,5}{1}{1}
%\epsilon_{5612b}\epsilon_{6123a}
%%\label{line6e} 
%\nonumber
\\
&-
\So\,
\BGraph{1,3,5}{2,4}{1,3,5}{1}{1}  
 H( a , b, c ) 
%\left[
%\frac{\epsilon_{1245a}
%      (\epsilon_{1245b}c_3-\epsilon_{12345}(b\cdot c))}{(u_2-1)d_{14}d_{25}}
%-2d_{35}a_1c_2b_4-2d_{13}a_5c_4b_2
%\right]
%\nonumber
%\\
%&
\quad -
\So\,
\BGraph{1,3,5}{2,4,6}{1,5}{1}{1}
%\left[
%-\frac{ (a\cdot b) \epsilon_{12645}\epsilon_{b1245}}{(u_2-1)d_{14}d_{25}}
%+b_4 f_{L}(a,b)  +b_2 f_R(a,b)
%\right]
 K( a , b) 
\nonumber
\\
&+
\frac{\Se}2\,
\BGraph{1,3,5}{2,4,6}{1,5}{1}{1}
\left[
 \frac{\epsilon_{1354a}\epsilon_{1352b}\epsilon_{4561b}}
     {\epsilon_{12345}d_{46}} + \frac{\epsilon_{1352a}\epsilon_{1354b}\epsilon_{5612b}}
     {\epsilon_{12345}d_{26}} 
\right]
\nonumber
\\
&-
\frac{\Se}{12}\,
\BGraph{1,3,5}{2,4,6}{1,3,5}{1}{1}
\epsilon_{135ba}\epsilon_{135bc}.
\label{eqn:B3}
\end{align}
%\endgroup
%\end{widetext}
where we have introduced the shorthand notation $d_{ij}=(i\cdot j)$, $a_i=(a\cdot i)$, $b_i=(b\cdot i)$, $c_i=(c\cdot i)$, and $\epsilon_{ijklm}=\epsilon(i,j,k,l,m)$ for the five-dimensional $\epsilon$ tensor. 
The numerators are expressed in terms of products of
$\epsilon$-tensors, making local finiteness manifest.
For example, the product of two tensors $\epsilon_{a,i,i+1,i+2,i_3}\epsilon_{b,i+3,i+4,i+5,i,b}$ suppresses all collinear divergences associated with loop $a$ and $b$, since 
\begin{align}
\epsilon_{a,i,i+1,i+2,i+3}\epsilon_{b,i+3,i+4,i+5,i} \rightarrow 0 \,,  \quad  \text{as} \quad   a \rightarrow  t_a y_i + (1-t_a) y_{i+1} , \;  b \rightarrow   t_b y_i + (1-t_b) y_{i+1}, \; \forall i \,. \nonumber
\end{align} 
The seventh and eighth finite integrals are constructed as
linear combinations of divergent integrals associated with
their subtopologies. Their numerators are given explicitly by
\begin{align}
K(a,b)={}&
\frac{(a\cdot b)\epsilon_{12456}\epsilon_{1245b}}
     {(u_2-1)d_{14}d_{25}}
+\frac{\epsilon_{56123}\epsilon_{5623b}\,a_1b_4}
      {(u_3-1)d_{13}d_{25}d_{36}}
+\frac{\epsilon_{34561}\epsilon_{3461b}\,a_5b_2}
      {(u_1-1)d_{14}d_{35}d_{36}}
\nonumber\\
&+2d_{15}(a\cdot b)(b_2d_{46}+b_4d_{26})
-2d_{15}\bigl(a_6b_2b_4+a_4b_2b_6+a_2b_4b_6\bigr).
\end{align}
and
\begin{align}
H(a,b,c) = 
\frac{
( (a\cdot b)  \epsilon_{12345} - a_3 \epsilon_{1245b} ) \epsilon_{1245c}
}{
(u_2-1)d_{14}d_{25}
}
+2d_{35}c_1a_2b_4
+2d_{13}c_5a_4b_2.
\end{align}

\iffalse
\begin{align}
&
 f_L (a,b)=  \frac{\epsilon_{26ab\star}\epsilon_{5613\star}}{d_{36}}
+d_{15}a_2b_6-d_{15}d_{26}(a\cdot b) , 
\\
&  
f_R(a,b)=
\frac{\epsilon_{46ab\star}\epsilon_{1653\star}}{d_{36}}
+d_{15}a_4b_6-d_{15}d_{46}(a\cdot b).
\end{align}
The numerator of each term in \eqref{eqn:B3} is expressed in terms of
$\epsilon$-tensors, making its local finiteness manifest. 
This representation in terms of simple, finite integrands enables integration in Feynman parameter space term by term using spherical contour projection. 
\fi

%\newpage

The parity-odd integrand $I^{\rm odd}$ likewise consists of ten locally finite terms, represented by bipartite graphs spanning the top topology and seven sub-topologies:
\begin{align}
I^{\rm odd} 
& ={} 
%\frac{d_{24}d_{46}d_{26}d_{13}d_{35}d_{15}}
%     {3d_{14}d_{25}d_{36}}
   \frac{2\So\Se}{3}
     \,
\BGraph{1,3,5}{2,4,6}{1,3,5}{1}{0}
\epsilon_{135ab}
%\label{line0ao}
\nonumber  
\\
&
-
\frac14\,
\BGraph{1,3,5}{2,4,6}{1,3,5}{1}{1}
\epsilon_{135ab}
\left[
\frac{\epsilon_{23456}\epsilon_{1234c}\epsilon_{4561b}}
     {\epsilon_{12345}d_{46}} + \frac{\epsilon_{12346}\epsilon_{2345c}\epsilon_{5612b}}
     {\epsilon_{12345}d_{26}} 
\right]
\nonumber
%\label{line0bo}
\\
&+
\frac{d_{13}}2\,
\BGraph{1,3}{2,4}{1,3}{1}{1}
\epsilon_{1234b}
%\label{line3oc}
%\\
%&
\quad +
\frac{d_{15}}2\,
\BGraph{1,5}{2,4,6}{1,5}{1}{1}
b_1\epsilon_{2456b}
\nonumber 
%\label{line4oc}
\\
&-
\frac{d_{13}d_{35}d_{15}}6\,
\BGraph{1,3,5}{2,4,6}{1,3,5}{0}{1}
\epsilon_{246bc}
%\label{line8oc}
%\\
%&
\quad 
+
d_{35}\,
\BGraph{1,3,5}{2,4,6}{1,3,5}{0}{1}
\left[
\frac{\epsilon_{12456}\epsilon_{1256b}
      d_{13}d_{24}c_3}
     {\epsilon_{12345}d_{26}}
-\epsilon_{1234c}d_{15}b_6
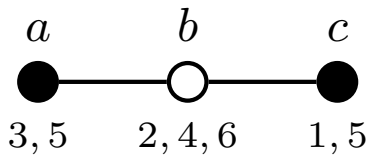
\right] 
\nonumber
%\label{line5oc}
\\
&+
\BGraph{3,5}{2,4,6}{1,5}{1}{1}
\left[
%\begin{aligned}
%&
\frac{\epsilon_{1234b}\epsilon_{1456b}\epsilon_{23456}}
      {2 d_{13}d_{24}d_{46}}
%\\
%&
-\frac{\epsilon_{12456}\epsilon_{1456b}
        d_{24}d_{35}b_3}
       {\epsilon_{12345}d_{46}}
 -\epsilon_{2345b}d_{15}b_6
%\end{aligned}
\right] 
%\label{line6oc}
\nonumber
\\
&+
\frac{d_{15}}2\,
\BGraph{1,3,5}{2,4}{1,3,5}{1}{1}
%\Bigl[
%%\begin{aligned}
%%&
%-\epsilon_{2345a}b_1c_3
 %+\epsilon_{1345a}b_2c_3
%%\\
%%&
%+d_{35}
% \bigl[
 %\epsilon_{124ab}c_3
 %+\epsilon_{1234c}(a\cdot b)
 %\bigr]
 \, G_{\rm o} (a,b,c)
%\end{aligned}
%\Bigr]
%\label{line7oc}
%\nonumber
%\\
%&
\quad +
\frac12\,
\BGraph{1,3,5}{2,4}{1,3,5}{1}{1}
\, H(\epsilon_{135a\star},b, c)
%\left[
%%\begin{aligned}
%%&
%\frac{\epsilon_{135ab}\epsilon_{12345}
 %      \epsilon_{1245c} }{(u_2 -1) d_{14} d_{25}}
%%\\
%%&
%+ 2 \epsilon_{135a2}c_1d_{35}b_4 + 2 \epsilon_{135a4}c_5d_{13}b_2 
%%\end{aligned}
%\right] 
\nonumber
\\
&
 +
\frac34\,
\BGraph{1,3,5}{2,4,6}{1,5}{1}{1}
\, K(\epsilon_{135a\star},b)
%\begin{aligned}
%&\epsilon_{135ab}
% \left[
% \frac{\epsilon_{12456}\epsilon_{1245b}}{(u_2 -1) d_{14} d_{25}}
% +2d_{15}(b_2d_{46}+d_{26}b_4)
% \right] \nonumber
%\\
%&-2d_{15}
% \bigl[
% \epsilon_{135a6}b_2b_4
% +\epsilon_{135a4}b_2b_6
% +\epsilon_{135a2}b_4b_6
% \bigr]
%\end{aligned}
%\right].  
%\label{line8oc}
\end{align}
where $H(\epsilon_{135a {\star}} ,b, c) $ and $K(\epsilon_{135a\star}, b) $ are obtained from $H(a,b,c)$ and $K(a,b)$, respectively,
 by replacing $a_{\mu}$ with $\epsilon(1, 3,5, a, {}_\mu)$.
 The numerator of the 8th term is: 
\begin{align}
G_{\rm o} (a,b,c) &= 
%\begin{aligned}
%&
-\epsilon_{2345a}b_1c_3
 +\epsilon_{1345a}b_2c_3
%\\
%&
+d_{35}
 \bigl[
 \epsilon_{124ab}c_3
 +\epsilon_{1234c}(a\cdot b)
 \bigr]
%\end{aligned}
\end{align}
Its integral vanishes, as discussed below.

\subsection{integrated results}
\label{app:integration}

% Keep footnotes in one full-width column in this one-column appendix.
\makeatletter
\let\set@footnotewidth\set@footnotewidth@one
\let\compose@footnotes\compose@footnotes@one
\onecolumn@grid@setup
\count\footins=1000\relax
\makeatother

In this appendix, we present the integrated results for each of the finite
integrals appearing in $I^{\rm even}$ and $I^{\rm odd}$, which can provide
valuable input to the computation of higher-order amplitudes. The integration is
performed in Feynman parameter space using the method of spherical contour
projection~\cite{Arkani-Hamed:2017ahv}. This provides an independent check of
the result presented in Sec.~\ref{sec:integration} of the main text.
All integrals are reduced to deformed
scalar integrals, with individually divergent contributions kept in finite
combinations. Local finiteness ensures that the integration can be carried
out without introducing additional regulators.\footnote{Tensor reduction
produces two types of contributions: bulk integrals over the deformation
parameters and boundary terms at infinite values of these parameters.
These contributions may diverge separately, but their divergences cancel
in the sum when a common boundary limit is taken.}

The integrated results for the ten integrals appearing in $I^{\rm even}$
are given explicitly below.

% First term of (B1).
\begin{equation*}
\mathcal I_1=
\frac56\Se\,d_{13}d_{35}d_{15}\int_{a,b,c}\!
\BGraph{1,3,5}{2,4,6}{1,3,5}{0}{0}
=\frac{5\pi^3}{48}\sigma_{246}.
\end{equation*}

% Second term of (B1), summed over the three cyclic images.
\begingroup
\setlength{\multlinegap}{0pt}
\begin{multline*}
\mathcal I_2+{\rm cyclic}\times2=
\frac{\So}{2}\int_{a,b,c}\!
\BGraph{1,3,5}{2,4,6}{1,3,5}{0}{1}
\left[
\frac{\epsilon_{23456}\epsilon_{1234c}\epsilon_{4561b}}
     {\epsilon_{12345}d_{46}}
+\frac{\epsilon_{12346}\epsilon_{2345c}\epsilon_{5612b}}
     {\epsilon_{12345}d_{26}}
+{\rm cyclic}\times2
\right]
\\
=\frac{\pi\sigma_{135}}{8}\left[
-6\bigl(\arccos\sqrt{u_1}\bigr)^2
-\log(4u_2)\log(4u_3)
+\left(1-\frac23\eta\right)\pi^2
+\operatorname{cyc}(u_1,u_2,u_3)
\right].
\end{multline*}
\endgroup

% Third term of (B1).
\begin{equation*}
\mathcal I_3=
\frac{\Se}{4d_{26}d_{46}d_{24}(u_2^{-1}-1)}\int_{a,b,c}\!
\BGraph{1,5}{2,4}{1,5}{1}{1}
\epsilon_{1245b}\epsilon_{12456}
=0.
\end{equation*}

% Fourth term of (B1).
\begin{equation*}
\mathcal I_4=
\frac{\Se}{4d_{26}d_{46}}\int_{a,b,c}\!
\BGraph{1,5}{2,4,6}{1,5}{1}{1}
\epsilon_{4561b}\epsilon_{5612b}
=\frac{\pi^3}{16}\sigma_{246}.
\end{equation*}

% Fifth term of (B1).
\begingroup
\setlength{\multlinegap}{0pt}
\begin{multline*}
\mathcal I_5=
\frac{\Se}{2d_{24}d_{46}}\int_{a,b,c}\!
\BGraph{1,3}{2,4,6}{1,5}{1}{1}
\epsilon_{1234b}\epsilon_{4561b}
=\frac{\eta\pi\sigma_{246}}{4}\widetilde I_{\rm critter}(3)
\\
=\frac{\eta\pi\sigma_{246}}{8}\left[
\frac{2\pi^2}{3}-\operatorname{Li}_2(1-u_1)-\operatorname{Li}_2(1-u_2)
-\log u_1\log u_2-2\bigl(\arccos\sqrt{u_3}\bigr)^2
\right].
\end{multline*}
\endgroup

% Sixth term of (B1).
\begin{equation*}
\mathcal I_6=
\frac{\So}{2d_{13}d_{25}}\int_{a,b,c}\!
\BGraph{1,3,5}{2}{1,3,5}{1}{1}
\left[\epsilon_{1352a}\epsilon_{1352c}
+2d_{13}d_{35}d_{15}a_2c_2\right]
=\frac{\pi^3}{24}\sigma_{135}.
\end{equation*}

% Seventh term of (B1).
\begin{equation*}
\mathcal I_7=
-\So\int_{a,b,c}\!
\BGraph{1,3,5}{2,4}{1,3,5}{1}{1}
H(a,b,c)
=-\frac{\pi^3}{4}\sigma_{135}.
\end{equation*}

% Eighth term of (B1).
\begingroup
\setlength{\multlinegap}{0pt}
\begin{multline*}
\mathcal I_8=
-\So\int_{a,b,c}\!
\BGraph{1,3,5}{2,4,6}{1,5}{1}{1}
K(a,b)
=\frac{\pi^3\sigma_{135}}{4}
-\frac{\pi\sigma_{135}}{2}\widetilde I_{\rm critter}(2)
\\
=\frac{\pi\sigma_{135}}{4}\left[
\operatorname{Li}_2(1-u_1)+\operatorname{Li}_2(1-u_3)
+\log u_1\log u_3
+2\bigl(\arccos\sqrt{u_2}\bigr)^2+\frac{\pi^2}{3}
\right].
\end{multline*}
\endgroup

% Ninth term of (B1).
\begingroup
\setlength{\multlinegap}{0pt}
\begin{multline*}
\mathcal I_9=
\frac{\Se}{2}\int_{a,b,c}\!
\BGraph{1,3,5}{2,4,6}{1,5}{1}{1}
\left[
\frac{\epsilon_{1354a}\epsilon_{1352b}\epsilon_{4561b}}
     {\epsilon_{12345}d_{46}}
+\frac{\epsilon_{1352a}\epsilon_{1354b}\epsilon_{5612b}}
     {\epsilon_{12345}d_{26}}
\right]
\\
=\frac{\eta\pi\sigma_{246}}8\left[
2\operatorname{Li}_2(1-u_2^{-1})+\log^2(4u_2)
+4\bigl(\arccos\sqrt{u_2}\bigr)^2\right].
\end{multline*}
\endgroup

% Tenth term of (B1).
\begin{equation*}
\mathcal I_{10}=
-\frac{\Se}{12}\int_{a,b,c}\!
\BGraph{1,3,5}{2,4,6}{1,3,5}{1}{1}
\epsilon_{135ba}\epsilon_{135bc}
=0.
\end{equation*}

In these expressions, we use the shorthand notation
\begin{equation*}
\sigma_{135}=\frac{\So}{\sqrt{d_{13}d_{15}d_{35}}},
\qquad
\sigma_{246}=\frac{\Se}{\sqrt{d_{24}d_{26}d_{46}}},
\qquad
\eta=-\sigma_{135}\sigma_{246},
\qquad
\sigma_{246}=\eta\sigma_{135}.
\end{equation*}
Here $\widetilde I_{\rm critter}(i)$ denotes the finite, dual-conformal
combination of the two-loop critter integral and its associated
double-triangle integral, defined in Ref.~\cite{Caron-Huot:2012sos}.
The index $i$ labels its cyclic images.

Summing over cyclic permutations, we observe cancellations among
$\mathcal I_2$, $\mathcal I_5$, $\mathcal I_8$, and $\mathcal I_9$, so that
the polylogarithms drop out of the sum, yielding a remarkably simple result
that agrees with the parity-even calculation in
Eqs.~\eqref{eq:collapse} and~\eqref{eq:one-loop}:

\begin{equation*}
\boxed{
\int_{a,b,c}I^{\rm even}+(i\mapsto i+2k)
=\frac{\pi^3}{4}(\sigma_{135}+\sigma_{246})
}.
\end{equation*}

For the odd-part integrals, all except two terms vanish upon integration.
In particular, the eighth term carrying the numerator $G_{\rm o}$ vanishes
by symmetry. The $\epsilon$-tensor identities give
$G_{\rm o}+G_{\rm o}^{\rm flip}=a_3c_3\,\epsilon_{1245b}$, so its integrand
can be written as a term odd under the dihedral flip
$(1\leftrightarrow5,\,2\leftrightarrow4)$ plus a term whose integral vanishes.
The integrated result can depend only on $u_2$, which is invariant under
this flip, and therefore must vanish.
The two non-vanishing contributions, the sixth and seventh terms of (B4),
are given below.

% Sixth term of (B4).
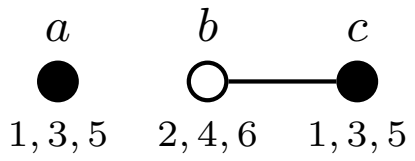
\begin{equation*}
\adjustbox{max width=\textwidth}{$\displaystyle
d_{35}\int_{a,b,c}\!
\BGraph{1,3,5}{2,4,6}{1,3,5}{0}{1}\left[
\frac{\epsilon_{12456}\epsilon_{1256b}d_{13}d_{24}c_3}
 {\epsilon_{12345}d_{26}}-\epsilon_{1234c}d_{15}b_6\right]
=\frac{\pi\omega}{4}\log\frac{v_2}{v_1}\log\frac{u_3}{u_1}.
$}
\end{equation*}

% Seventh term of (B4).
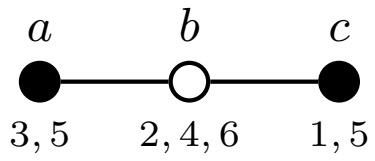
\begin{equation*}
\adjustbox{max width=\textwidth}{$\displaystyle
\int_{a,b,c}\!
\BGraph{3,5}{2,4,6}{1,5}{1}{1}\left[
\frac{\epsilon_{1234b}\epsilon_{1456b}\epsilon_{23456}}{2d_{13}d_{24}d_{46}}
-\frac{\epsilon_{12456}\epsilon_{1456b}d_{24}d_{35}b_3}
 {\epsilon_{12345}d_{46}}
-\epsilon_{2345b}d_{15}b_6\right]
=\frac{\pi\omega}{4}\log v_1\log\frac{u_3}{u_2}.
$}
\end{equation*}

Here the variables $v_1$ and $v_2$ parametrize the cross-ratios, and the
prefactor $\omega$ is defined as follows:

\begin{gather*}
u_1=\frac{(1+v_1)^2}{4v_1},\qquad
u_2=\frac{(v_1+v_2)^2}{4v_1v_2},\qquad
u_3=\frac{(1+v_2)^2}{4v_2},
\\
\omega=\frac{\epsilon_{12345}\sqrt{v_1v_2}}
 {(v_1-v_2)\sqrt{d_{13}d_{14}d_{24}d_{25}d_{35}}},
\qquad \omega^2=-\frac12.
\end{gather*}

The cyclic sums of these two non-vanishing contributions cancel, giving
\begin{equation*}
\boxed{
\int_{a,b,c}I^{\rm odd}+(i\mapsto i+2k)=0
}.
\end{equation*}

\end{document}